\documentclass[preprint,showpacs,amsmath,amssymb,aps,prc,superscriptaddress]{revtex4}
\usepackage{epsfig}

\begin{document}

\preprint{APS/123-QED}

\title{\bf \large Cluster Model For Reactions Induced By Weakly Bound And/Or 
Exotic Halo Nuclei With Medium-Mass Targets}

\bigskip

\author{C. BECK\thanks{Correponding author: 
christian.beck@ires.in2p3.fr},$^{1,*}$
N. ROWLEY,$^{1,+}$ 
P. PAPKA,$^{1,++}$ 
S. COURTIN,$^{1}$ 
M. ROUSSEAU 
}

\affiliation{\it Institut Pluridisciplinaire Hubert Curien - D\'epartement de
Recherches Subatomiques, UMR7178, 
IN2P3-CNRS et Universit\'{e} de Strasbourg, 23 rue du Loess, 
B.P. 28, F-67037 Strasbourg Cedex 2, France} 

\medskip

\author{
F.A. SOUZA,$^{ 2}$ 
N. CARLIN,$^{ 2}$ 
F. LIGUORI NETO,$^{2}$ 
M.M. DE MOURA,$^{2}$ 
M.G. DEL SANTO,$^{ 2}$ 
A.A.I. SUADE,$^{ 2}$ 
M.G. MUNHOZ,$^{ 2}$ 
E.M. SZANTO,$^{ 2}$ 
A. SZANTO DE TOLEDO 
}

\medskip

\address{Departamento de Fisica Nuclear, Universidade de S\~ao Paulo, SP, 
Brazil}

\author{N. KEELEY}

\address{The Andrzej Soltan Institute for Nuclear Studies, Warsaw, Poland}

\author{A. DIAZ-TORRES} 

\address{Department of Physics, University of Surrey, Guildford, Surrey, GU2 
7XH, UK}

\author{K. HAGINO}

\address{Department of Physics, Tohoku University, Sendai, Japan}

\date{\today}

\newpage

\begin{abstract}

{An experimental overview of reactions induced by the stable, but weakly-bound 
nuclei $^6$Li, $^7$Li and $^9$Be, and by the exotic, halo nuclei $^6$He,
$^{8}$He, $^8$B, and $^{11}$Be on medium-mass targets, such as $^{58}$Ni, 
$^{59}$Co or $^{64}$Zn, is presented. Existing data on elastic scattering, 
total reaction cross sections, fusion processes, breakup and transfer channels 
are discussed in the framework of a CDCC approach taking into account the 
breakup degree of freedom.} 

\end{abstract}

\maketitle

\noindent
$^{+}$ Institut Physique Nucl\'eaire, Orsay, France
$^{++}$ Department of Physics, University of Stellenbosch, 7602 Matieland, 
South Africa

\newpage

\section{Introduction}

In reactions induced by weakly bound nuclei and/or by halo nuclei, the 
influence on the fusion process of coupling both to collective degrees of 
freedom and to transfer/breakup channels is a key point for the understanding 
of N-body systems in quantum dynamics~[1]. Due to their very weak binding 
energies, a diffuse cloud of neutrons for $^{6}$He or an extended spatial 
distribution for the loosely bound proton in $^{8}$B would lead to larger 
total reaction (and fusion) cross sections at sub-barrier energies as compared 
to predictions of one-dimensional barrier penetration models~[1,2]. This
enhancement is well understood in terms of the dynamical processes arising 
from strong couplings to collective inelastic excitations of the target 
(such as "normal" quadrupole and octupole modes) and projectile (such as 
soft dipole resonances). However, in the case of reactions where at least 
one of the colliding nuclei has a sufficiently low binding energy for 
breakup to become a competitive process, conflicting conclusions were 
reported~[1-4]. Recent studies with Radioactive Ion Beams (RIB) indicate that 
the halo nature of $^{6}$He, for instance, does not enhance the fusion 
probability as anticipated. Rather the prominent role of one- and 
two-neutron transfers in $^{6,8}$He induced fusion reactions~[3] was 
definitively demonstrated. On the other hand, the effect of non-conventional 
transfer/stripping processes appears to be less significant for stable 
weakly bound projectiles~[5-15]. Several experiments involving $^{9}$Be, 
$^{7}$Li, and $^{6}$Li projectiles on medium-mass targets have been
undertaken~[1,5-15]. In this contribution, a comprehensive study of the 
$^{6,7}$Li+$^{59}$Co reactions~[5-11] is presented as a benchmark as
illustrated by Fig.~1. 

\section{Experimental results.}

The fusion excitation functions as measured for the $^{6,7}$Li+$^{59}$Co 
reactions [5] are presented in Fig.~1 with comparisons with other lighter 
targets [12-15].

\begin{figure}[th]
\centerline{\psfig{figure=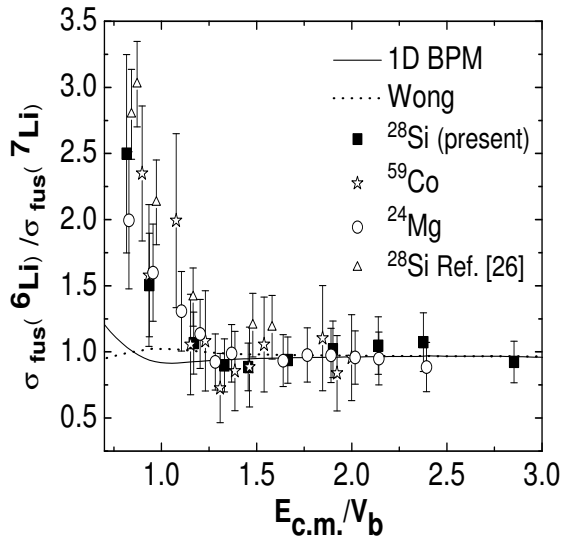,width=8.3cm,height=8.3cm}}
\vspace*{8pt}
\caption{\label{fig1} Ratios of measured fusion cross sections for $^{6}$Li
and $^{7}$Li projectiles with $^{24}$Mg [12], $^{28}$Si [13,14] and $^{59}$Co 
[5,6] targets as a function of E$_{c.m.}$/V$_b$. The solid line gives the 
1D-BPM prediction [5] while the dotted line shows results obtained from Wong's 
prescription [5]. This figure taken from Refs.[14,15] was originally shown
for $^{6,7}$Li+$^{59}$Co in Ref.[5].}
\end{figure}

A comparison with Continuum-Discretized Coupled-Channel (CDCC) calculations
[6,7,11] indicates only a small enhancement of total fusion for the more weakly 
bound $^{6}$Li below the Coulomb barrier, with similar cross sections for both 
$^{6,7}$Li+$^{59}$Co reactions at and above the barrier. It is interesting to
notice that the same conclusions have been found for both $^{24}$Mg [12] and 
$^{28}$Si [13-15] targets (see Fig.~1). This result is consistent with rather 
low breakup cross sections measured for the $^{6,7}$Li+$^{59}$Co reactions 
even at incident energies larger than the Coulomb barrier [8-10]. However, the 
coupling of the breakup channel is extremely important for the CDCC analysis 
[7-11] of the angular distributions of the elastic scattering. Therefore, a 
more detailed investigation of the breakup process in the $^{6}$Li+$^{59}$Co 
reaction with particle coincince techniques has also been proposed to discuss 
the interplay of fusion and breakup processes [8-11]. Coincidence data 
compared to three-body kinematics calculations [9,10] reveal a way how to 
disentangle the contributions of breakup, incomplete fusion and/or 
transfer-reemission processes. 

\begin{figure}[th]
\centerline{\psfig{figure=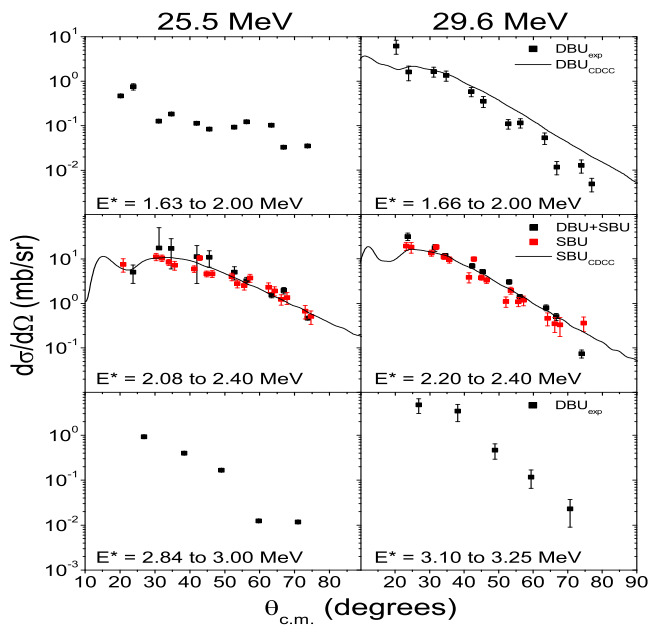,width=9.6cm,height=9.0cm}}
\vspace*{8pt}
\caption{\label{fig2} Experimental and theoretical CDCC angular distributions
for the SBU [8] and DBU [9-11] projectile breakup processes (see text for 
details) obtained at E$_{lab}$ = 25.5 MeV and 29.6 MeV [10] for 
$^{6}$Li+$^{59}$Co. The chosen experimental continuum excitation energy ranges 
are given. }
\end{figure}

\section{Discussion and conclusions}

Fig.~2 displays experimental (full rectangles) and theoretical angular 
distributions (solid lines) for the sequential (SBU) and direct (DBU) 
projectile breakup processes [8-10] at the two indicated bombarding energies 
for the $^{6}$Li+$^{59}$Co reaction. In the CDCC calculations [7,11] the 
$\alpha$ + $d$ binning scheme has been appropriately altered to accord exactly 
with the measured continuum excitation energy ranges. For this reaction it has 
not been necessary to use the sophisticated four-body CDCC framework proposed 
by M. Rodriguez-Gallardo in this conference [16]. The CDCC cross sections are 
in agreement with the experimental ones, both in shapes and magnitudes within 
the uncertainties.
The relative contributions of the $^{6}$Li SBU and DBU to the incomplete
fusion/transfer process has been discussed in Refs.[9-11] by considering
the corresponding lifetimes obtained by using a semi-classical approach
[9,10]. We concluded that the flux diverted from complete fusion to incomplete
fusion would arise essentially from DBU processes via high-lying continuum
(non-resonant) states of $^{6}$Li; this is due to the fact that both the
SBU mechanism and the low-lying DBU processes from low-lying resonant
$^{6}$Li states occur at large internuclear distances [10]. Work is in
progress to study incomplete fusion for $^{6}$Li+$^{59}$Co within a newly
developed 3-dimensional classical trajectory model [17,18]. 

As far as exotic halo projectiles are concerned we have initiated a systematic
study of $^{8}$B and $^{7}$Be induced reactions~[2] with an improved CDCC 
method [7]. As compared to $^{7}$Be+$^{58}$Ni (similar to 
$^{6,7}$Li+$^{58,64}$Ni) the CDCC analysis of $^{8}$B+$^{58}$Ni reaction
while exhibiting a large breakup cross section (consistent with the
systematics [19,20]) is rather surprizing as regards the consequent 
weak coupling effect found to be particularly small on the near-barrier 
elastic scattering [2]. A full understanding of the reaction dynamics 
involving couplings to the breakup and nucleon-transfer channels will need 
high-intensity RIB and precise measurements of elastic scattering, fusion 
and yields leading to the breakup itself.

\begin{center}

\end{center}

\end{document}